# Universal Fabrication of Graphene/Perovskite Oxide Hybrid Heterostructures

Yeongju Choi[#, 1], Seungjin Lee[#, 2], Dongwon Shin[1, 6], Sukhoon Sim[1], Min-Hyoung Jung[2], Dirk Wulferding[3], Minjae Kim[1], Jaesik Eom[1], Myeesha Mostafa[5, 6], Wonhee Ko[5, 6], SeungNam Cha[1], Jungseek Hwang[1], Hu Young Jeong[4], Ki Kang Kim*[, 2] and Woo Seok Choi*[, 1]

*[1]Department of Physics, Sungkyunkwan University, Suwon 16419, Republic of Korea*

*[2]Department of Energy Science, Sungkyunkwan University, Suwon 16419, Republic of Korea*

*[3]Department of Physics and Astronomy, Sejong University, Seoul 05006, Republic of Korea*

*[4]Graduate School of Semiconductor Materials and Devices Engineering, Ulsan National Institute of Science and Technology, Ulsan 44919, Republic of Korea*

*[5]Department of Physics and Astronomy, The University of Tennessee, Knoxville, Tennessee 37996, United States*

*[6]Center for Advanced Materials and Manufacturing, The University of Tennessee, Knoxville, Tennessee 37996, United States*

* Author to whom correspondence should be addressed:

kikangkim@skku.edu (Ki Kang Kim) and choiws@skku.edu (Woo Seok Choi)

[#] These authors contributed equally to this work.

**Abstract**

Hybrid heterostructures composed of graphene and perovskite oxides provide a promising platform for exploiting synergetic interfacial functionalities. Conventional fabrication methods of the hybrid heterostructures rely on transferring graphene grown on metallic substrates— a process that is time-consuming, labor-intensive, and prone to introducing numerous defects. In this study, we present a universal, catalyst-free method for the direct growth of graphene on insulating substrates by employing three different perovskite oxide substrates ($SrTiO_3$, $LaAlO_3$, and $(La_{0.18}Sr_{0.82})(Al_{0.59}Ta_{0.41})O_3$) using atmospheric chemical vapor deposition. Comprehensive characterization via Raman spectroscopy, X-ray spectroscopy, scanning probe microscopy, and electron microscopy confirmed the formation of a uniform, continuous monolayer graphene on all substrates. We identified that growth temperature critically governs graphene quality, as excessive active species may lead to secondary nucleation and the formation of multilayer graphene. Notably, all substrates shared the same optimal growth conditions. Low-temperature Raman spectroscopy and scanning tunneling microscopy of the graphene/$SrTiO_3$ hybrid heterostructure revealed cooperative phenomena, including substrate-induced lattice-phonon and electron-phonon coupling. Our work establishes a reproducible, transfer-free fabrication route for graphene/perovskite oxide hybrid heterostructures and provides empirical support for the universal growth of graphene on insulating substrates.

# 1. Introduction

Hybrid heterostructures (HHs) composed of graphene and perovskite oxides serve as an emerging platform for exploiting interface-engineered cooperative functionalities.[1, 2] The platform hosts synergetic phenomena originating from the interaction between the two distinct materials, realizing properties that are not observable in conventional heterostructures.[3-5] A prominent example is the HH composed of the most well-studied 2D material, graphene, and the perovskite oxide $SrTiO_3$ (STO) which form one of the most celebrated heterointerfaces owing to its unique electronic interactions. Graphene (Gr)/STO HHs exhibit virtually noninteracting Dirac electrons,[6] substantial voltage scaling of the quantum Hall effect, unexpected magnetoresistance, and a helical quantum Hall phase.[7] The graphene layer can also serve as a probe for investigating the light-modulated surface polarization and oxygen-vacancy-modulated dielectric responses of STO.[8] With the emergence of potential remote epitaxy, Gr/STO HHs have also been utilized to fabricate free-standing perovskite oxide membranes.[9-12] However, with few exceptions, most studies have relied on graphene transfer to fabricate the HHs. This is because graphene typically grows preferentially on metallic substrates such as Cu or Ni, where catalytic activity facilitates chemical vapor deposition (CVD), unless exfoliated.[13-16] Although growth on metallic substrates generally yields high-quality graphene, the subsequent transfer process often introduces unintended defects and results in low yield. Moreover, the process is time-consuming, involves complex manual handling, and is limited in the achievable lateral size.

To overcome these challenges, the direct growth of graphene on perovskite oxide substrates can be envisaged. Conventionally, direct graphene growth has been demonstrated on catalytically active metal substrates. However, achieving direct growth of 2D layers including graphene on insulating substrates remains challenging.[17] On metals, catalytic activity facilitates the decomposition of carbon precursors, which promotes graphene formation.[13-16, 18] In contrast, insulating substrates lack such activity, resulting in different growth mechanisms.[19] Consequently, graphene growth on these substrates tends to have a narrow growth window and a low growth rate.[20-26] In the absence of catalysis, thermal decomposition is expected to dominate, suggesting that graphene growth may be relatively insensitive to specific insulating substrate used. Therefore, assuming the substrates remain stable under growth conditions, establishing universal growth conditions for HH fabrication may be possible.

Given that graphene growth on insulating substrates might be relatively independent of substrate choice, we herein demonstrate the universal fabrication of graphene/perovskite oxide HHs using a simple atmospheric pressure chemical vapor deposition (APCVD) process. We used (001)-oriented STO, LaAlO$_3$ (LAO), and (La$_{0.18}$Sr$_{0.82}$)(Al$_{0.59}$Ta$_{0.41}$)O$_3$ (LSAT) perovskite oxide substrates, as they are the widely used and thermally stable up to at least 1100 °C. Ar, H$_2$, and CH$_4$ gases were used as precursors in the APCVD process, which was conducted at relatively high temperatures (> 1050 °C) to enable thermal decomposition. Upon identifying a narrow but optimal growth window, we achieved uniform graphene growth across the entire 5 × 5 mm$^2$ substrate area. The key factors for high-quality graphene growth were precise control of temperature and precursor flow rate. Most importantly, we identified a set of growth conditions that were universally applicable to STO, LAO, and LSAT, thereby potentially eliminating the need for transfer process in HH fabrications.

## 2. Results and Discussion

### 2.1 HH fabrication

We identified an optimal growth condition applicable to Gr/STO,[9, 23] Gr/LAO, and Gr/LSAT HHs. We controlled temperature, precursor flow rate, and growth time, in the APCVD process. The optimal conditions were found to be a temperature of 1070 °C, a mixed gas of Ar, $H_2$, and $CH_4$ with a flow rate ratio of 250 : 10 : 8 sccm, and a growth time of 1–3 hours (see Methods section). In this study, $CH_4$ was selected as the carbon source to ensure consistency with established precedents for catalyst-free graphene growth on insulating substrates.[9, 20-25] While $C_2H_6$ (ethane) is a possible alternative, its higher reactivity is anticipated to promote unintended carbon aggregation or graphene adlayers.[27] The overall quality of the HHs was evaluated by assessing the physical characteristics of graphene, uniformity, and cleanliness of the surfaces and interfaces.

## 2.2 Raman and X-ray Spectroscopies

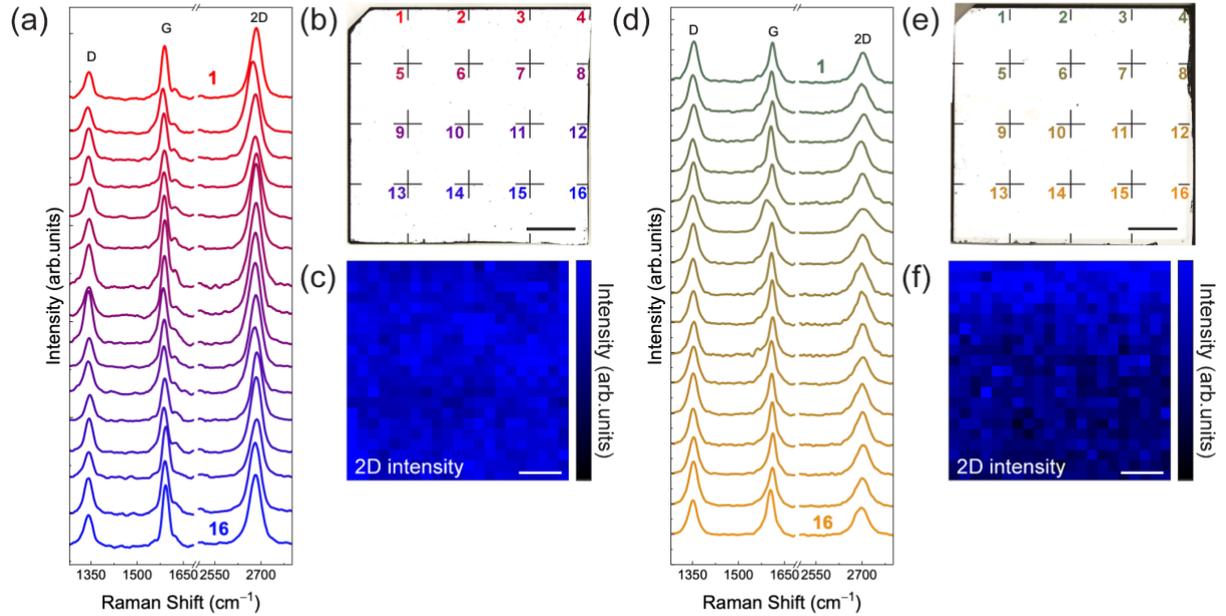

**Figure. 1 Uniform graphene growth on (a, b, c) STO and (d, e, f) LAO substrates.** (a, d) Raman spectra from 16 evenly spaced points on the HH surfaces, demonstrating graphene uniformity. The positions of the measured points are marked as different colors and numbers on the optical microscopy images in (b, e) (scale bar: 1 mm). (c, f) 2D intensity mappings of Raman spectra from the HHs taken at random positions (scale bar: 4 μm).

Under optimal growth conditions, the HHs exhibit uniform graphene quality. The Raman spectra in **Figure 1** demonstrate the uniformity of the Gr/STO and Gr/LAO HHs.[28] We subtracted the substrate spectra from the HH spectra to eliminate the substrate contributions, using confocal Raman spectroscopy (Figure S1, Supporting Information).[29] Figure 1a and 1d show the Raman spectra measured at equal lateral intervals (as indicated in Figure 1b and 1e), across the 5 × 5 mm$^2$ Gr/STO and Gr/LAO HHs, respectively. Figure 1b and 1e are the optical microscope images of the grown graphene samples; the corresponding images without numbers or boundary lines for better visualizations are shown in Figure S2. Furthermore, the statistical analysis presents in Figure S3 which plots the average and standard deviation of the spectra from Figure 1a and 1d, displays narrow deviation bands, confirming the uniformity across the HH samples. Raman spectroscopy for the Gr/LSAT HH is shown in Figure S4a. Although a strong fluorescence feature at high energy from the LSAT substrate (Figure S1c, Supporting Information) obscured part of the analysis, similar uniformity was confirmed (Figure S4a, Supporting Information). All Raman spectra across the substrates exhibited the characteristic D, G, and 2D peaks of graphene. In Figure 1c and 1f, we mapped the 2D peak intensity to visualize the spatial uniformity of the Gr/STO and Gr/LAO HHs. The 2D peak intensity was nearly uniform across the HHs, with the spatial resolution is limited by the laser spot size (~600 nm). The G peak intensity maps also indicated uniformity across the samples (Figure S5, Supporting Information). We note that Raman spectra of graphene may vary across different substrates, possibly due to dielectric screening and strain effects originating from differences in thermal expansion coefficients (TECs).[30-32] Therefore, although a 2D/G intensity ratio below two is typically taken to indicate polycrystalline nature or lower graphene quality on conventional SiO$_2$/Si substrates, this criterion may not apply directly to Gr/perovskite oxide HHs.[23] Nevertheless, our Raman spectroscopy results clearly demonstrate that the graphene was uniformly grown across all perovskite substrates employed.

Similarly, near-edge X-ray absorption fine structure (NEXAFS) spectra of the HHs confirmed the quality of the graphene layer.[33, 34] C $K$-edge NEXAFS spectra reveal clear features of $\sigma^*$ C=C bonding at 292 eV and $\pi^*$ C-C bonding at 286 eV, as shown in Figure S6. Considering that the measurement was performed vertically, the relatively sharp feature and higher intensity of $\sigma^*$ peak compared to those of $\pi^*$ peak indicate the growth of graphene, rather than other phases, such as graphene oxide. The peak features originating from graphene were nearly identical regardless of the substrate used, indicating a similar quality of graphene.

## 2.3 Scanning and Electron Microscopies

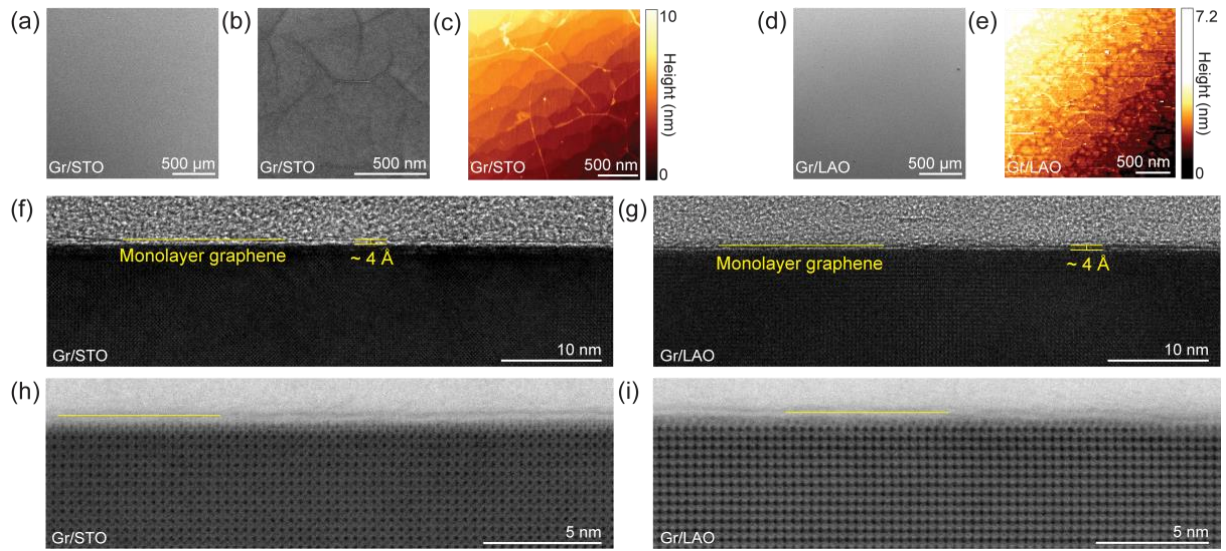

**Figure. 2 Micrometer- and atomic-scale microscopy images of HHs with (a–c, f, h) STO and (d, e, g, i) LAO substrates.** (a, b, d) SEM images and (c, e) AFM topography images of HH surfaces. The dark lines in the SEM image (b) and the bright lines in the AFM images (c, e) represent graphene wrinkles. Cross-sectional (f, g) TEM images and (h, i) STEM images of Gr/STO and Gr/LAO HHs, respectively. Dark lines on the crystal structures indicate monolayer graphene (highlighted with yellow horizontal lines).

Micrometer- and atomic-scale microscopy confirm the high structural quality of the HH surfaces and interfaces. Scanning electron microscopy (SEM) revealed clean surfaces at the micrometer scale under optimal growth conditions (**Figure 2**a, 2b, 2d). When the growth conditions deviated from the optimum, defects, residues, or overgrowth readily appeared on the surface (Figure S7, Supporting Information). For the Gr/STO HH, high-magnification SEM imaging (Figure 2b) was possible because the STO substrate became metallic with the introduction of oxygen vacancies under the reducing, high-temperature conditions of APCVD.[35] At higher magnification, we observed several lines on the surface, previously interpreted in the literature as graphene wrinkles or grain boundaries.[36-39] Atomic force microscopy (AFM) images revealed similar line structures atop the step-and-terrace structures of the substrates, as shown in Figure 2c and 2e, for Gr/STO and Gr/LAO HHs, respectively. The surfaces of Gr/STO and Gr/LAO HHs were atomically flat, with RMS roughness values of 0.319 and 0.274 nm, respectively (Figure S8, Supporting Information), aside from the inherent step-and-terrace structures (~0.4 nm) of the substrates (Figure S9, Supporting Information). The characteristic lines exhibited heights of ~1.2 and ~0.6 nm for the Gr/STO and Gr/LAO HHs, respectively. The Gr/LSAT HH also exhibited similar surface quality and the line features in AFM (Figure S8c, d, g, j, Supporting Information), although an extensive number of particulates originating from the annealed substrate were observed (Figure S9c, Supporting Information).[40] Additionally, lateral force microscopy (LFM) (Figure S10, Supporting Information) revealed similar frictional distributions, and conductive AFM (C-AFM) (Figure S11, Supporting Information) demonstrated uniform conductivity across the HHs, both supporting the high surface quality of the graphene.

Cross-sectional transmission electron microscopy (TEM) and scanning TEM (STEM) confirmed predominant monolayer graphene growth in the HHs. The dark horizontal lines observed atop of the perovskite oxide surface in all TEM (Figure 2f, 2g, and Figure S4b, Supporting Information) and STEM images (Figure 2h, 2i, and  Figure S4c, Supporting Information) indicate successful formation of monolayer graphene on STO, LAO, and LSAT substrates.[27, 41] The distance between graphene and the top perovskite oxide surface was approximately 4 Å with small variation, consistent with previous reports on graphene grown on $SiO_2$ and Cu substrates.[42, 43] Multilayer graphene was rarely observed only in regions containing extended surface defects (Figure S12, Supporting Information), suggesting that a self-limiting mechanism governs graphene formation on clean perovskite oxide surfaces.

We further identified graphene wrinkles, which appeared as characteristic line structures in the SEM and AFM images (Figure 2b, c, e). Cross-sectional TEM and STEM of the Gr/STO HH showed that the wrinkles had a height of ~1.4 nm (Figure S13, Supporting Information), consistent with the AFM observations (Figure S8e, Supporting Information). The images indicate that the monolayer graphene remains continuous even in the presence of wrinkles, with no evident cracks or disruptions. We speculate that these wrinkles result from the TEC mismatch between graphene and the substrates.[44-46] During the post-growth cooling process, graphene expands while the perovskite oxide substrate contracts, leading to natural wrinkle formation. While such feature may raise concerns about the quality of the graphene surface, the TEC-induced wrinkle is a common phenomenon, even observed in high-quality, uniform graphene grown on Cu foils.[47] Overall, microscopic observations including SEM, AFM, LFM, C-AFM, TEM, and STEM, confirm the successful growth of high-quality monolayer graphene on perovskite oxide substrates under optimized conditions.

## 2.4 Temperature Dependent Growth

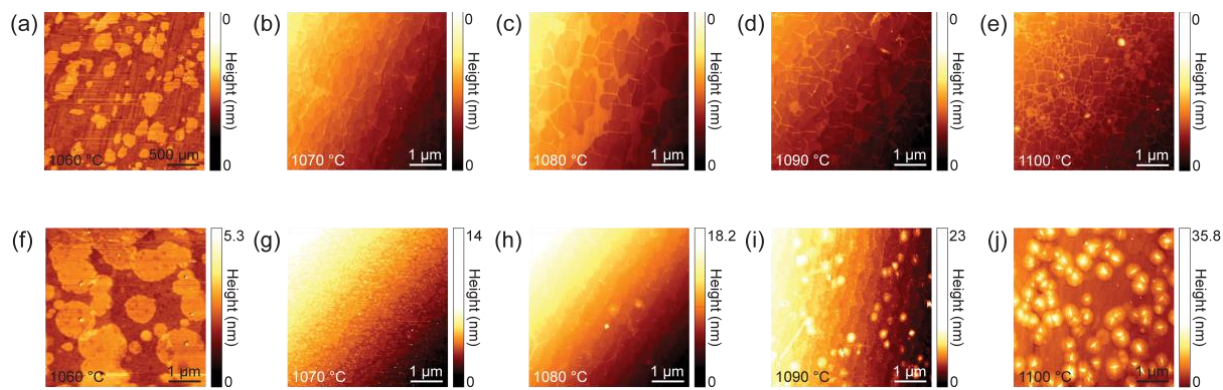

**Figure. 3 Temperature-dependent growth behavior of the HHs.** AFM topography images of (a–e) Gr/STO and (f–j) Gr/LAO HHs. As the growth temperature increases, adlayer or secondary nucleation becomes more prominent.

In optimizing the growth conditions for HHs, temperature was found to be one of the most critical parameters, and the observed trends were consistent across all the substrates used in this study. We first noted that graphene began to grow above 1060 °C. As shown in **Figure 3**a, no uniform graphene features were observed at 1060 °C. At 1070 °C, graphene grew uniformly across the substrate, with visible wrinkles (Figure 3b). At temperatures above 1070 °C, adlayers (relatively bright regions) began forming atop the graphene surface (Figure 3c–e). This secondary nucleation is typically associated with reduced graphene quality when grown on dielectric substrates.[9, 24, 27, 48, 49] A similar trend was observed for graphene grown on LAO substrates, as shown in Figure 3f–j. Raman spectra of HHs fabricated at various temperatures corroborated the AFM results, with optimal spectra corresponding to samples grown at 1070 and 1080 °C (Figure S14, Supporting Information). We also observed that above 1070 °C, defect density and size increased with growth time (Figure S15, Supporting Information). However, at the optimal temperature of 1070 °C, a self-limiting mechanism seemed to suppress defect formation above the monolayer graphene, even with prolonged growth durations (Figure S16, Supporting Information). Other factors, such as hydrogen flow rate and substrate surface roughness, also significantly influenced the growth process (Supplementary Note 1; Figure S12 and S17, Supporting Information). Similar to temperature dependence, suboptimal values for these parameters promoted secondary nucleation, which hindered the formation of uniform, defect-free HHs. Therefore, suppressing secondary nucleation is key to achieving high-quality HH growth.

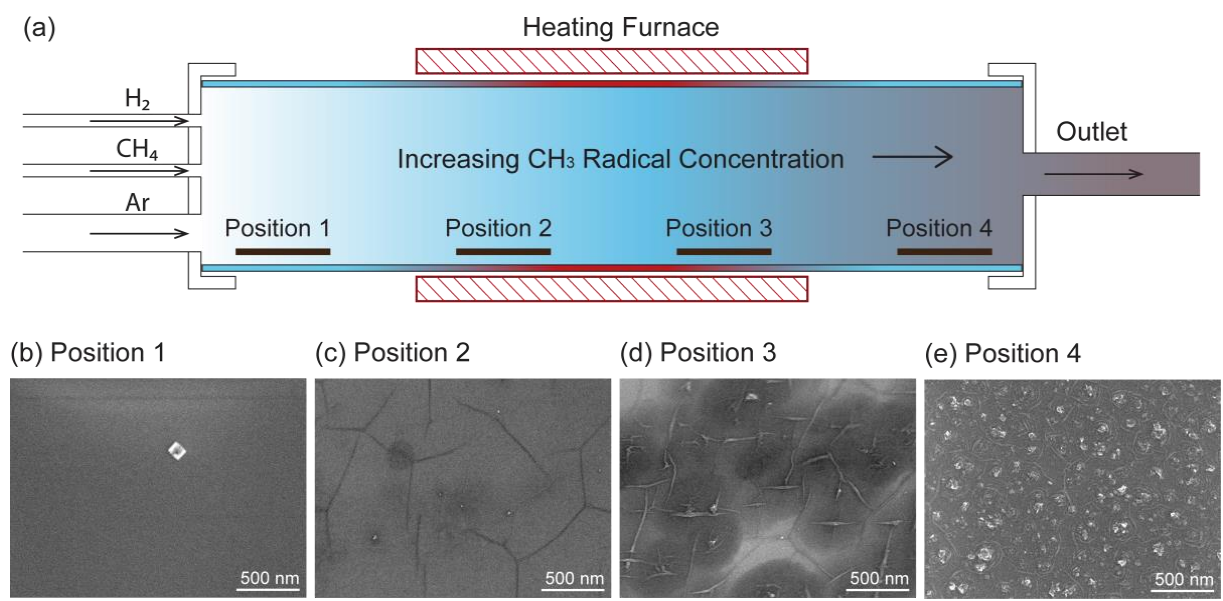

**Figure. 4 Position dependent growth behavior of Gr/STO HHs in APCVD at 1100 °C.** (a) Schematic representation of different growth positions within the APCVD quartz tube. The temperature peaks at the center of the tube, whereas $CH_3\cdot$ radical concentration is highest near the outlet (position (e)). (b–e) SEM images of HH surfaces corresponding to the Positions 1-4 in (a).

Secondary nucleation can result from oversupply of active species or surface reactions that promote overgrowth. The APCVD growth of graphene on insulating perovskite oxide surfaces can be described by a two-step mechanism: generation of active carbon precursors, followed by graphene formation from these precursors.[19, 50, 51] While purely thermal decomposition of $CH_4$ is unlikely at our growth temperature (1070 °C) on an insulating substrate,[50-52] we propose the decomposition maybe driven by a gas-phase interaction. Consistent with a recent computational study,[19] this mechanism considers hydrogen radicals (H·), abundantly supplied in our APCVD environment, collide with $CH_4$ molecules to facilitate the generation of $CH_3$· radicals (H· + $CH_4$ → $CH_3$· + $H_2$). This reaction has a sufficiently low activation energy (45 – 60 kJ/mol) to be readily activated at our experimental temperature.[19] The first step is thus a gas-phase reaction dependent on the gas temperature, whereas the second step is a surface reaction primarily governed by the substrate temperature.

To elucidate this mechanism, we positioned substrates at various locations within the APCVD quartz tube at 1100 °C, as shown in **Figure 4**a. As the substrate is positioned near the inlet (Position 1), growth does not initiate, because the temperature is not sufficiently high. As the substrate is positioned toward the center of the tube (Position 2), an ideal growth condition is achieved. The center of the tube apparently experiences the highest temperature, which enhances the both the $CH_3$· generation and the surface reaction rate. Hence, as the substrate is positioned beyond the center of the tube (Position 3), overgrowth of graphene takes place. Although $CH_3$· radical generation is most active near the center, the flow direction causes $CH_3$· radicals to accumulate toward the outlet. SEM images showed that the sample placed closest to the outlet (Position 4) exhibited the most severe secondary nucleation, indicating that excessive $CH_3$· radicals promote overgrowth. To confirm the effect of radical oversupply, we increased the $CH_4$ precursor flow rate from the optimal 8 to 16 sccm. This adjustment led to the formation of graphene adlayers, as seen in SEM and AFM images (Figure S18, Supporting Information). The relatively low, optimal 8 sccm of $CH_4$ flow rate is a necessary condition.

Based on previous studies[9, 19, 24, 27, 48, 49] and our own experimental results, graphene growth on an insulating substrate has a relatively slow growth rate compared to that on a metallic substrate. Therefore, if the $CH_3\cdot$ radical supply is too high, it would overwhelm the slow surface kinetics, causing secondary nucleation and adlayer formation as demonstrated in Figure 4e and Figure S18. Consequently, a relatively low $CH_4$ partial pressure is required to balance the gas-phase precursor supply with the slow surface kinetics, enabling self-limiting process of monolayer graphene growth for the HH.

Finally, the temperature dependence of graphene growth was consistent across STO, LAO, and LSAT substrates, with no major differences observed among these insulating substrates.[20, 21, 25, 27] The observation that STO shares identical growth temperature windows and defect trends with LAO despite the presence of oxygen vacancies (Figure 3) indicates that these vacancies exert negligible influence on the growth mechanism. (Supplementary Note2) This consistency provides strong experimental validation for our proposed gas-phase mechanism, which, being independent of the specific oxide substrate chemistry, explains the similar growth behaviors. These findings highlight the universal applicability of APCVD for fabricating HHs and underscore the importance of controlling $CH_3\cdot$ radical concentration. Optimizing external parameters such as growth temperature and substrate position is essential for achieving high-quality HHs.

## 2.5 Low-Temperature Raman Spectroscopy and Scanning Tunneling Microscopy

A structural phase transition in STO is reflected in the phonon spectra of graphene, providing an example of the synergetic behavior. To probe the cooperative interfacial interactions, we measured low-temperature Raman spectra of the HH. The antiferrodistortive (AFD) transition of STO, which occurs at ~100 K depending on the oxygen stoichiometry of STO, results in a change in its TEC.[53-55] As a consequence, the strain exerted onto graphene is expected to vary. **Figure 5** shows the temperature-dependent positions of the G and 2D peaks of the Gr/STO HH. Both peaks shift to higher frequencies as the temperature decreases. Notably, the slope of the G peak shift changes sign at ~100 K. The 2D peak also exhibits an anomaly near that temperature, although less pronounced. Below the AFD transition, the TEC of STO exhibits minimal temperature dependence, which is also reflected in the Raman spectra of graphene. In contrast, Gr/LAO HH does not show any significant temperature-dependent behavior (Figure S19, Supporting Information), supporting that the observed Raman peak shifts in Gr/STO HH primarily arise from substrate-induced thermo-mechanical interactions in the HHs.

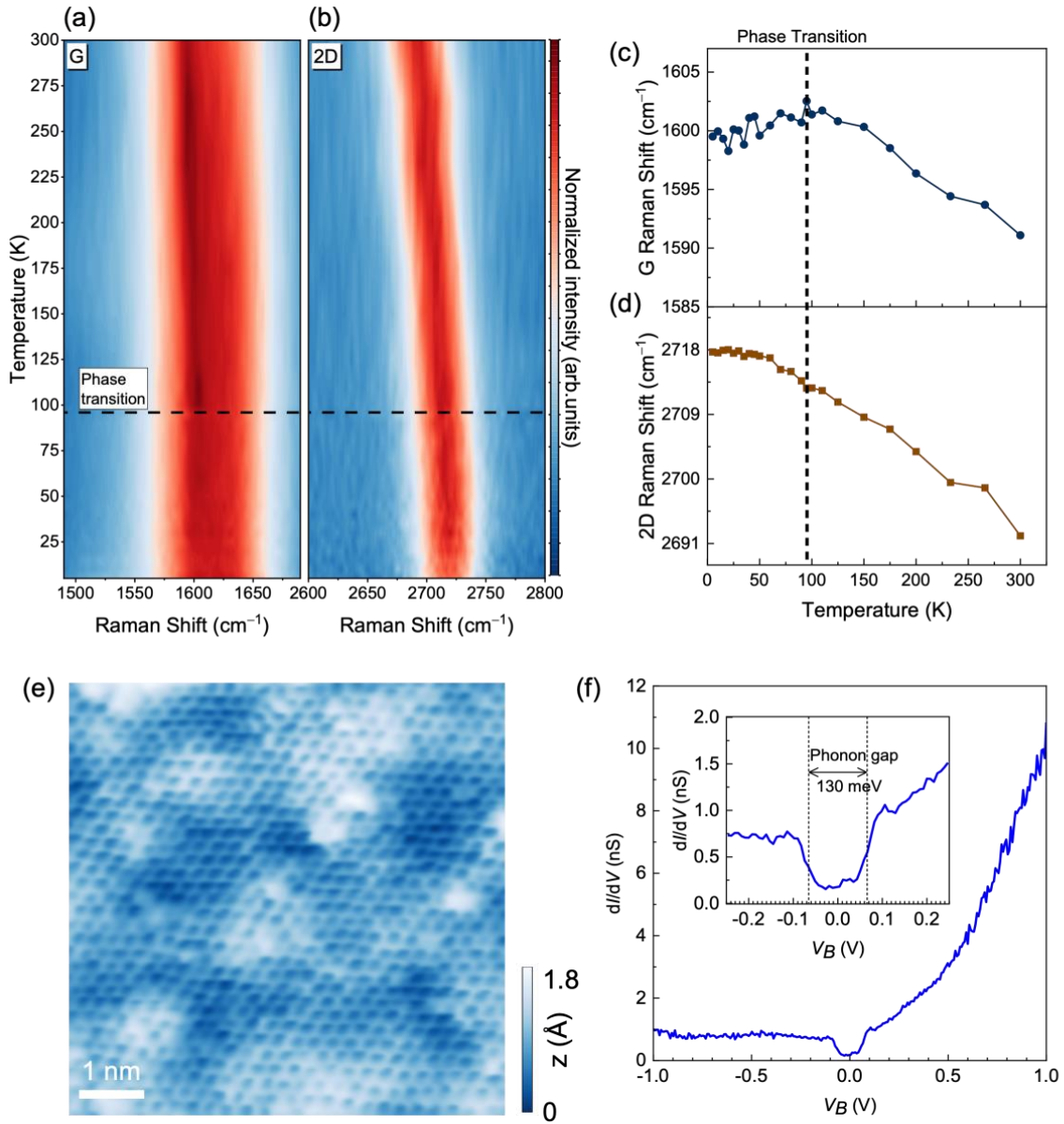

**Figure. 5 Low temperature Raman and STM/S Analyses of Gr/STO HH.** (a, b) Color contour plots of the temperature-dependent Raman spectra around the (a) G and (b) 2D peaks. AFD phase transition of STO occurs at approximately 100 K. (c, d) Peak positions of the G and 2D peaks as a function of temperature. (e) Topographic image of Gr/STO HH with atomic resolution ($V_B$ = 100 mV, $I_t$ = 1 nA) obtained by STM. (f) d$I$/d$V$ spectrum of Gr/STO HH obtained by STS. The inset shows the same spectra enlarged around the Fermi level.

To visualize the atomic lattice and investigate the electronic structure of the Gr/STO HH, we performed scanning tunneling microscopy/spectroscopy (STM/S) study at 4.2 K. Figure 5e presents a $5 \times 5$ nm$^2$ topographic image with atomic resolution, clearly resolving the honeycomb lattice of graphene. Figure 5f displays the characteristic differential conductance (d$I$/d$V$) spectrum of the surface. The spectrum shows strong asymmetry, with significantly reduced d$I$/d$V$ on the negative bias side compared to the positive side, indicating that the Dirac point of graphene lies below the Fermi energy owing to substantial electron doping induced by HH formation. Additionally, a gap-like feature appears at the Fermi level. The width of the gap is approximately 130 mV, consistent with the phonon-induced inelastic electron tunneling spectra.[56-58] This observation suggests that the local phononic structure of graphene is discernible by STS, providing another example of the synergetic behavior in the Gr/STO HH. While the measurement temperature of 4.2 K may be too high to resolve strain-induced shifts in phonon energy, future studies at lower temperatures would make such features observable.

## 3. Conclusion

In summary, we have demonstrated a universal, catalyst-free fabrication method for Gr/perovskite oxide (STO, LAO, and LSAT) HHs using an optimized APCVD process. Spectroscopic and microscopic characterization confirmed that all the HHs possessed uniform and continuous monolayer graphene. A systematic parameter scan revealed a narrow, yet robust growth window that governs the universal formation of optimal HHs. This further suggests that the same optimal conditions may be applicable to other insulating, high-temperature-stable substrates. In addition, our study goes beyond fabrication to provide a detailed investigation of the synergetic interfacial phenomena. Low-temperature spectroscopy revealed thermo-mechanical and electron-phonon interactions between graphene and the STO substrate, proving multiple evidence of the synergetic behavior in the HH. Our work establishes a transfer-free method for HH fabrication and offers empirical evidence for universal graphene growth principles on insulating substrates.

## 4. Experimental Section/Methods

*Pretreatment of Perovskite Oxide Substrates*: (001)-oriented STO, LAO, and LSAT substrates (Shinkosha, Japan) were selected as growth substrates. To obtain step-terrace structures on the surface, the substrates were annealed under AP conditions at temperatures ranging from 950 to 1250 °C, depending on the substrate type. [40, 59] Prior to annealing, STO and LAO substrates were chemically etched using a buffered oxide etchant for 15 seconds to selectively remove surface contamination and the native amorphous oxide layer, thereby promoting the formation of atomically smooth surface and well-defined step edges during subsequent annealing.

*Growth of Graphene on Perovskite Oxides*: The pretreated substrates were loaded into an APCVD chamber, which was purged for 20 minutes using high-purity argon (Ar, > 99.9999%) and hydrogen ($H_2$, > 99.999%) gases at flow rates of 250 and 10 sccm, respectively, to eliminate residual air and prevent unwanted reactions. Then, the temperature was rapidly ramped up to 1070 °C within 10 min. The specified temperature refers to the furnace setpoint at the center of the chamber measured by an external thermocouple. The substrates were annealed at this temperature for 15 min to remove native oxides and carbonaceous impurities from their surfaces. After annealing, methane gas ($CH_4$, > 99.999%) was introduced at a flow rate of 8 sccm for 1–3 h to facilitate graphene growth. After the growth, the system was naturally cooled down to room temperature.

*Spectroscopy Measurements*: The graphene film was analyzed using confocal Raman spectroscopy (532 nm; XperRAM 200, NanoBase, Korea). As the Raman signals from the substrate overlapped with those from graphene, the substrate spectrum was subtracted from the total spectrum at each corresponding position.[29] To account for intensity differences, both spectra were normalized by their respective standard deviations. Low-temperature Raman spectra were obtained using a custom-built Raman system with a 532 nm continuous-wave laser. Near-edge X-ray absorption fine structure (NEXAFS) measurements were conducted at the C $K$-edge (0° incidence, 300 K) to probe graphene-specific features, performed at the 6A beamline of the Pohang Accelerator Laboratory (PAL), Korea.

*Microscopy Measurements*: The HH samples were characterized using AFM (XE7, Park Systems, Korea) to assess topography and lateral force for investigating graphene and substrate morphology. Current mapping was performed to evaluate local conductance. Surface morphology and structural characteristics of the as-grown graphene films were evaluated by field-emission scanning electron microscopy (FE-SEM; JSM-7600F, JEOL, Japan), where defects, adlayers, or graphene wrinkles produced variations in secondary electron emission. An optical microscope was used to identify measurement areas and to verify the uniformity of the Raman spectra across the HHs. Cross-sectional specimens were prepared using a dual-beam focused ion beam (FIB) system (FEI Helios NanoLab 450). The FIB sampling process was conducted at 30 kV (90 pA) and refined at 5 kV (56 pA) to minimize the formation of amorphous layers caused by beam damage. TEM and STEM were performed using a JEM-ARM200CF equipped with a double Cs corrector, operating at an accelerating voltage of 200 kV and under ultra-high vacuum conditions.

*Scanning Tunneling Microscopy/Spectroscopy*: STM/STS measurements were performed using a Unisoku USM-1300 STM operated at ultra-high vacuum (< 5 × $10^{-10}$ Torr) and 4.2 K. To clean the surface, the sample was annealed at 400 °C for 1 h inside the STM chamber prior to measurement. Mechanically ground PtIr wires were used as STM tips, which were further prepared on a Au(111) crystal by performing field emission, followed by gentle poking and pulsing on a clean surface. To acquire d$I$/d$V$ spectra, we used a conventional lock-in technique with a modulation amplitude of 1 mV and modulation frequency of 873 Hz.

**Acknowledgements**

**Funding**

This work was supported by the National Research Foundation of Korea (NRF) grants funded by the Korea government (MSIT) (Nos., 2021R1A2C2011340, RS-2023-00220471, RS-2023-00281671, RS-2022-R1A2C2011109, and RS-2022-NR068223). M.M. and D.S. were partially supported by the National Science Foundation Materials Research Science and Engineering Center program through the UT Knoxville Center for Advanced Materials and Manufacturing (DMR-2309083). D.W. acknowledges support from the faculty research fund of Sejong University in 2025.

The authors declare no competing financial interest.

# Supporting Information

**Supplementary Note 1: Graphene Growth Depending on Various Parameters.**

Under optimal growth conditions, a self-limiting process leads to the formation of a continuous monolayer graphene film across the substrate, which remains stable even with extended deposition times (Figure S15). This stability is attributed to the low carbon solubility of the substrate, which prevents the formation of additional graphene layers (adlayers) beneath the primary monolayer.[1]

<u>Supplementary Note 1.1: Effect of Hydrogen Gas Flow Rate</u>

During HH growth, large defects formed on the surface when the $H_2$ flow rate was not optimal. AFM topography images in Supplementary Figure S17 illustrate the effect of $H_2$ flow rate on Gr/STO HHs. At high $H_2$ flow rates (100 and 40 sccm; Figure S15a, b, respectively), the number of small particles increased, eventually covering the entire surface at 100 sccm. This behavior correlates with the Raman spectra (Figure S17f), where increased D peak intensity indicates higher defect density.[2] We speculate this is due to C–H bond formation, as increased $H_2$ flow limits carbon precursor mass transport and promotes secondary nucleation.[3] Conversely, under $H_2$-deficient conditions (5 and 0 sccm; Figure S17d,e, respectively), overgrowth was observed. In particular, without $H_2$, the substrate was completely covered with graphite-like foil, as seen in the optical image (Figure S17e). Similar overgrowth in the absence of $H_2$ has been reported on Cu foils,[4, 5] suggesting that the correct $H_2$ flow rate is crucial to suppress the formation of graphitic overgrowth on insulating substrates.

<u>Supplementary Note 1.2: Influence of Substrate Surface Purity</u>

Surface defects can promote multilayer graphene growth. Even under optimal HH conditions,

a few regions exhibited multilayer graphene, as revealed by SEM and STEM (Figure S12). SEM images (Figure S12c) show that areas near particulate contaminants (white dots), likely originating from the quartz tube during high-temperature APCVD,[6] are prone to multilayer nucleation. Correspondingly, STEM images (Figure S12d) show up to three layers of graphene, highlighting the impact of these particles. In the Gr/LSAT HHs, SrO mounds also act as unstable nucleation sites, leading to by-products (Figure S1c). Thus, a clean, atomically flat, and stable substrate surface under high-temperature conditions is essential for high-quality HH formation.

**Supplementary Note 2: Potential introduction of oxygen vacancies in SrTiO$_3$ substrates.**

We believe that STO, LSAT, and LAO substrates share the same growth condition for the graphene formation. For STO, formation of oxygen vacancies (OVs) may induce catalytic behavior. However, we could not observe any conducting behavior of the STO surface. As shown in Figure S20, we show NEXAFS measurements at the Ti $L$-edge and O $K$-edge of the "STO OV" HH sample and the "STO" HH sample. STO OV HH sample is the as-grown sample with potential introduction of OVs, whereas STO HH sample is an annealed one to recover the STO stoichiometry. In the Ti $L$-edge spectra (Figure S20a), both samples display identical spectral features, including no spectral weight shift or emergence of additional peaks characteristic of Ti$^{3+}$ states. Similarly, the O $K$-edge spectra (Figure S20b) exhibit virtually indistinguishable spectral features. Consequently, at least on the surface of STO seems to be rather inert against potential oxygen vacancy formation.

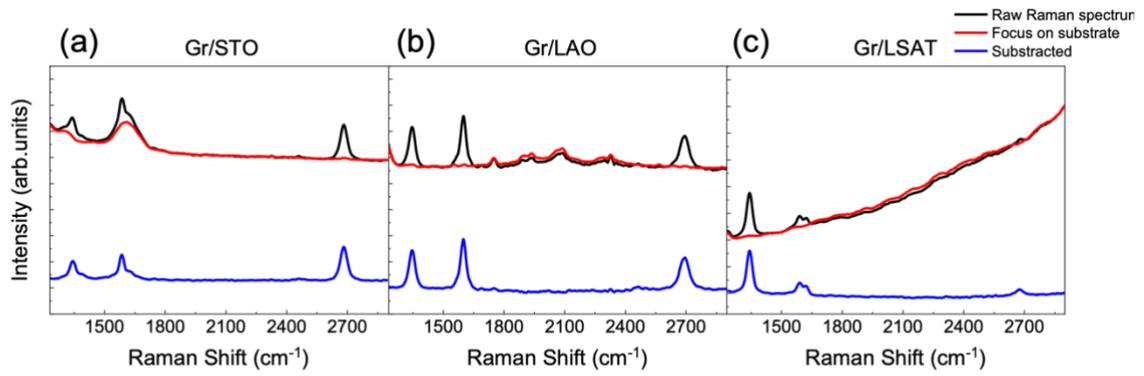

**Figure S1.** Subtraction of substrate Raman spectra from raw Raman spectra to isolate graphene features. Substrate spectra were collected at same locations as the raw spectra. (a) Gr/STO, (b) Gr/LAO, (c) Gr/LSAT HHs.

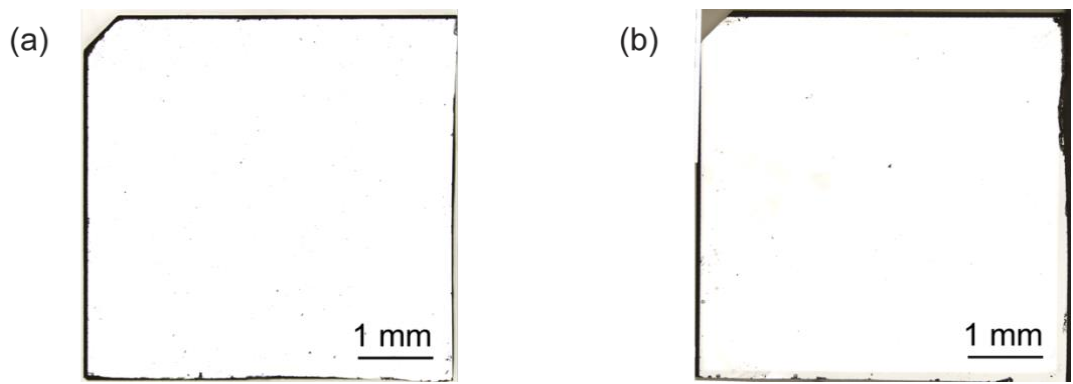

**Figure S2.** Large-area optical microscopy images of (a) Gr/STO and (b) Gr/LAO HHs. The images demonstrate the high uniformity and absence of large-scale defects, particulates, or delaminated regions across the $5 \times 5$ mm$^2$ sample surfaces.

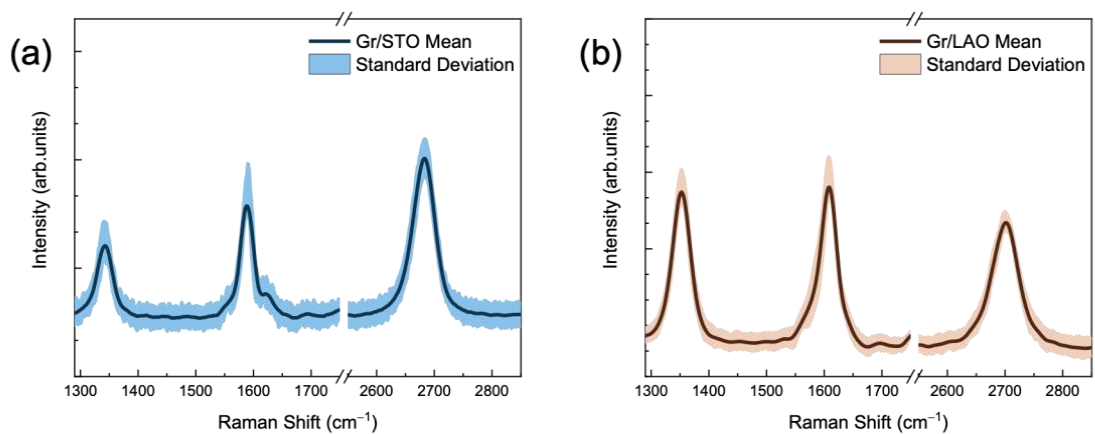

**Figure S3.** Statistical analyses of Raman spectra for (a) Gr/STO and (b) Gr/LAO HHs. The solid lines and the corresponding colored shaded regions represent the average and standard deviation of the Raman spectra on different positions of the HH.

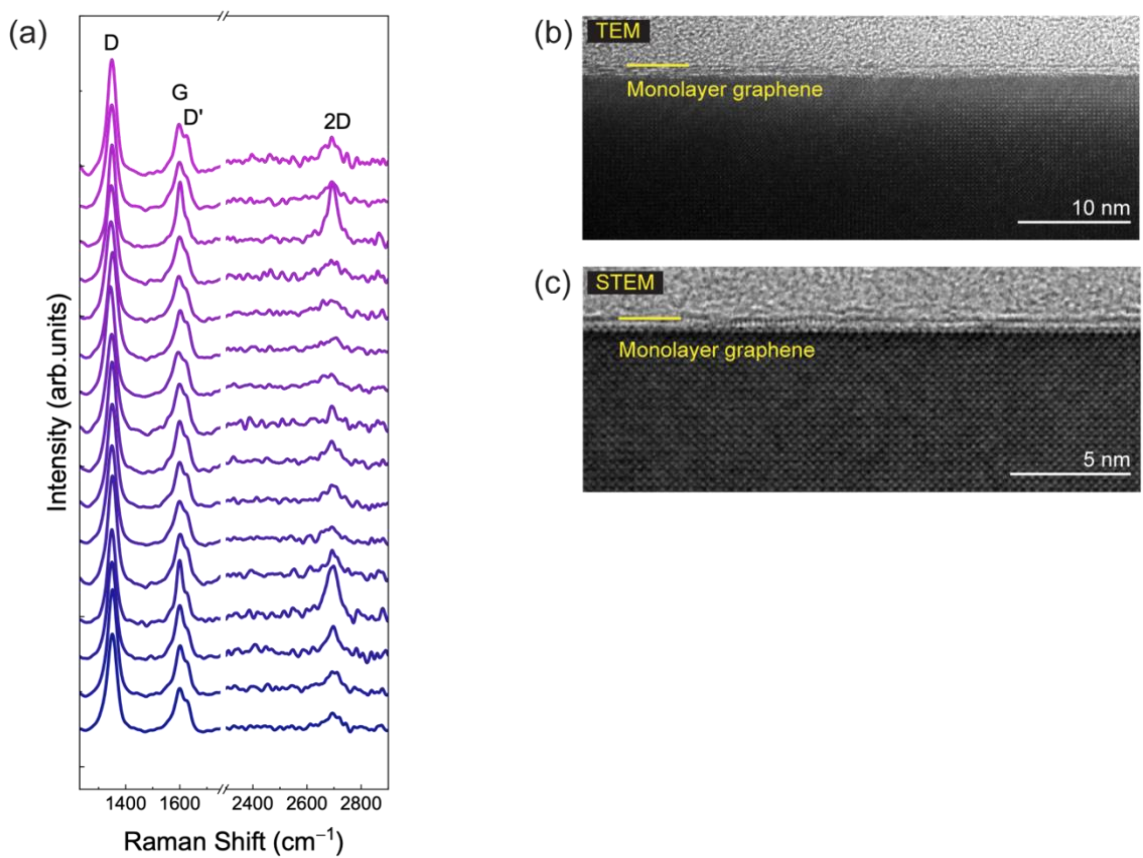

**Figure S4.** Raman spectra, TEM, and STEM characterization of Gr/LSAT HH. (a) Raman spectra from 16 evenly spaced points on the Gr/LSAT HH. (b) TEM and (c) STEM images of Gr/LSAT. Dark lines in both images (highlighted by horizontal yellow lines) indicate monolayer graphene.

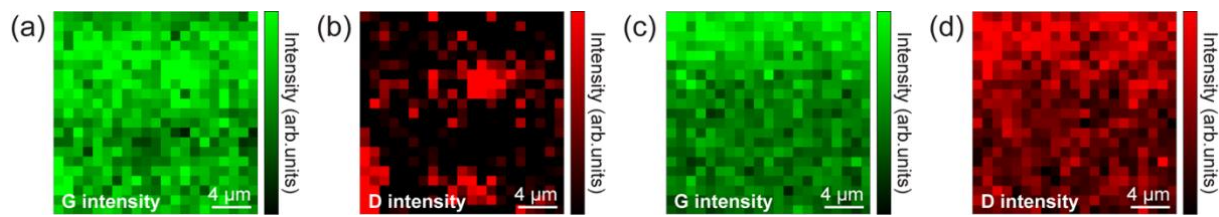

**Figure S5.** G and D peak intensity maps from Raman spectroscopy. (a, b) Gr/STO, (c, d) Gr/LAO. Uniform G peak intensity maps (a, c) confirm consistent graphene coverage.

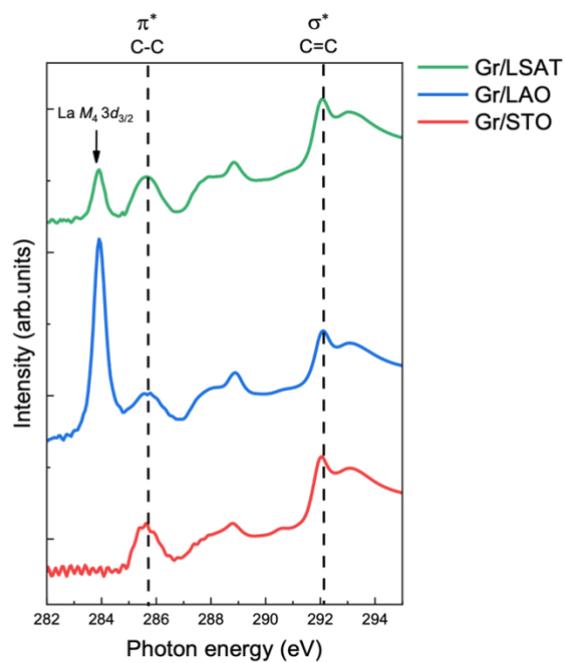

**Figure S6.** C *K*-edge NEXAFS spectra of the HHs. All samples exhibit σ* C=C bonding at 292 eV and π* C-C bonding at 286 eV of graphene. Broadening between these bands indicates potential interlayer or adsorbate-induced features.

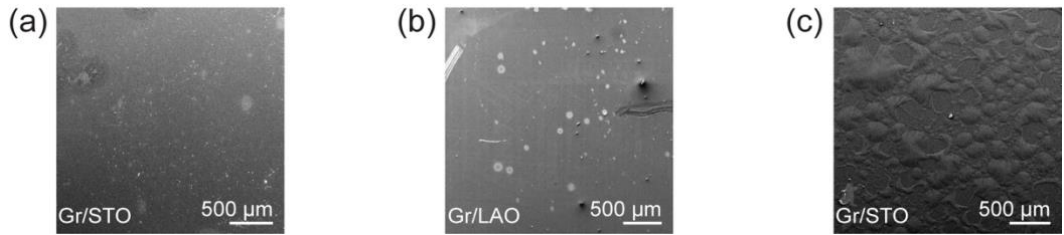

**Figure S7.** SEM images of growth in improper APCVD conditions. (a, b) Growth at high temperature and high precursor flow rate, showing defects as bright spots. (c) Overgrown graphene image when the substrate's growth position is near the outlet side of the APCVD chamber.

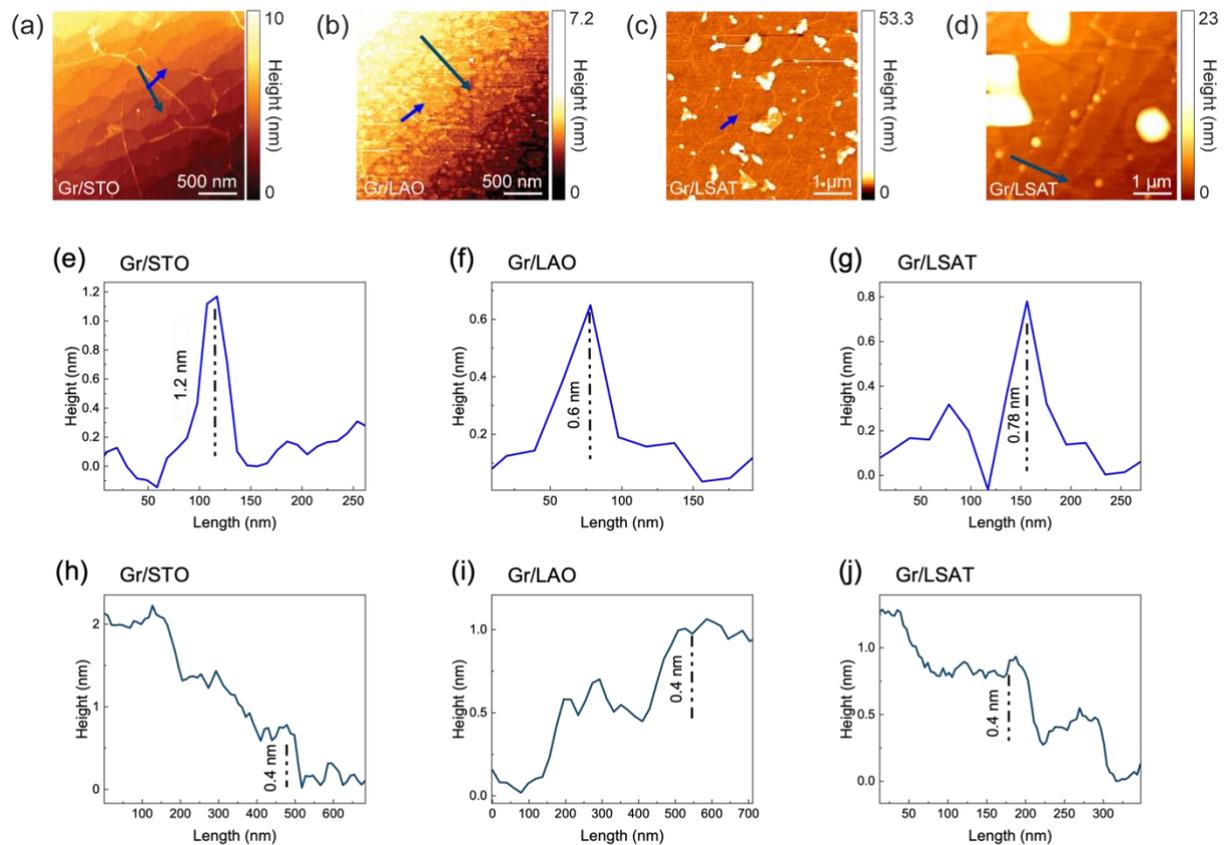

**Figure S8.** AFM analysis of graphene wrinkles and substrate steps. (a–d) Topography of Gr/STO, Gr/LAO, and Gr/LSAT HHs. (e–j) Line profiles along the arrows show comparable wrinkle (blue lines) (e–g) and step heights (navy lines) (h–j) for the HHs.

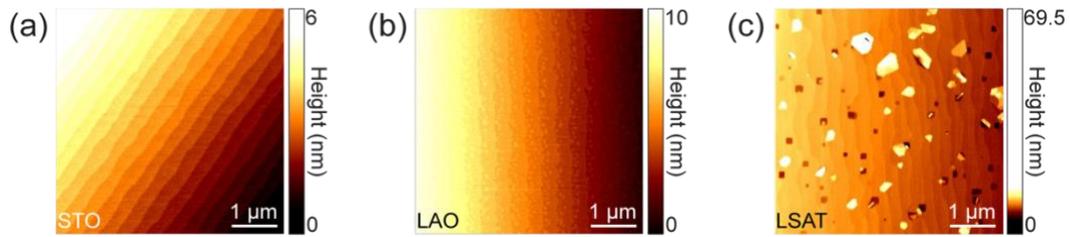

**Figure S9.** AFM topography images of pristine substrate: (a) STO and (b) LAO show flat step-terrace structures; (c) LSAT shows pits and mounds from surface SrO segregation.

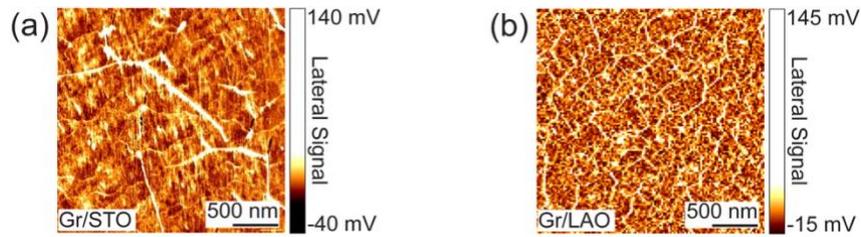

**Figure S10.** LFM images of HHs of Gr/STO (a) and Gr/ LAO (b). Increased friction observed at graphene wrinkles and substrate step edges.

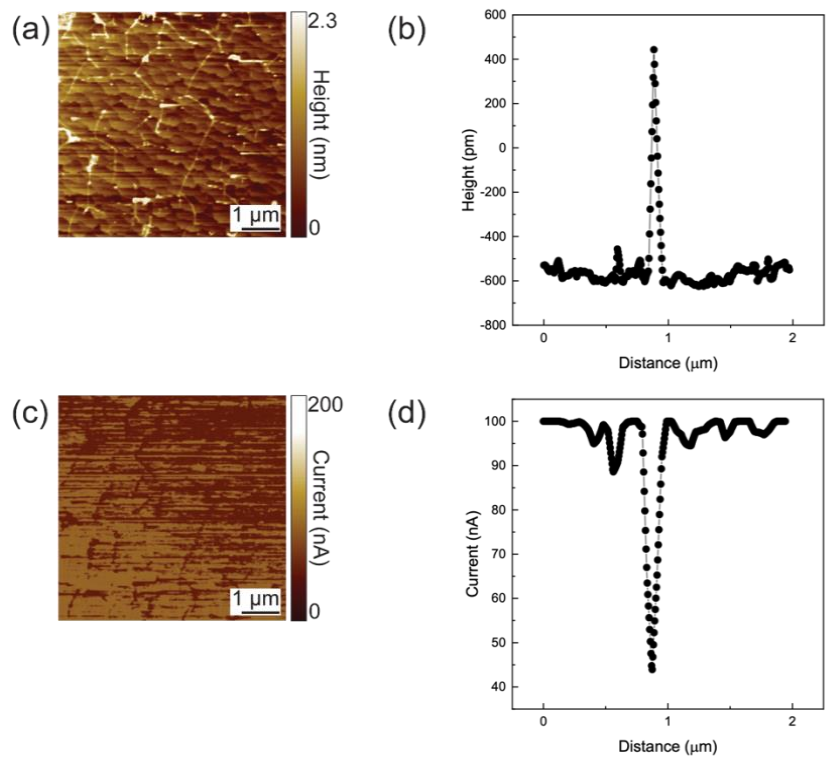

**Figure S11.** C-AFM images of Gr/STO. (a, b) AFM topography and (c, d) current maps with line profile from identical regions, revealing local conductivity variations.

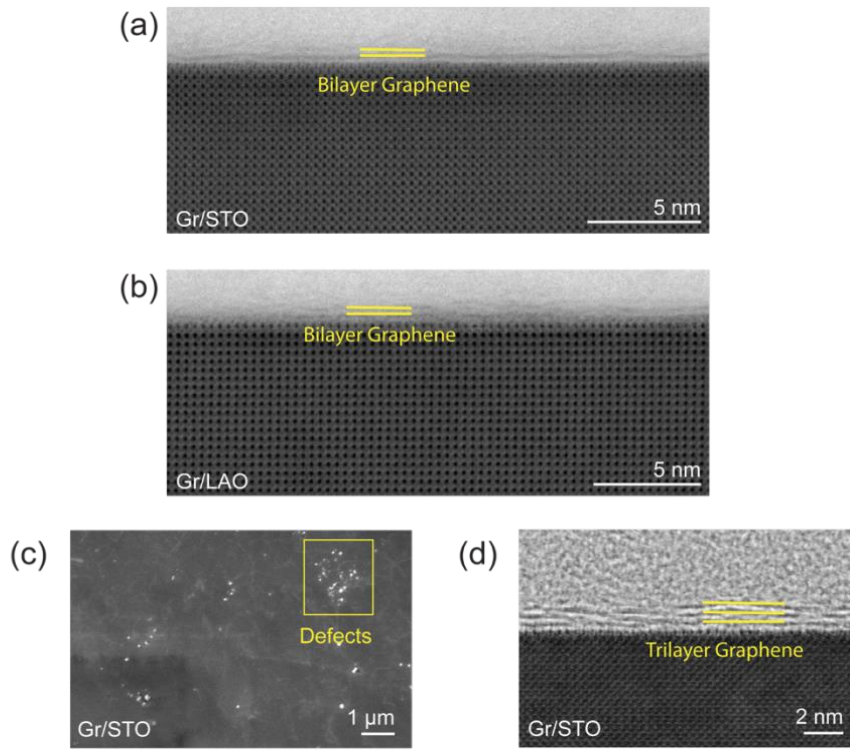

**Figure S12.** Cross-sectional STEM of multilayer (defective) regions. (a) Gr/STO, (b) Gr/LAO showing bilayer graphene (yellow lines). (c) SEM and (d) STEM near defects (yellow circle) showing trilayer graphene.

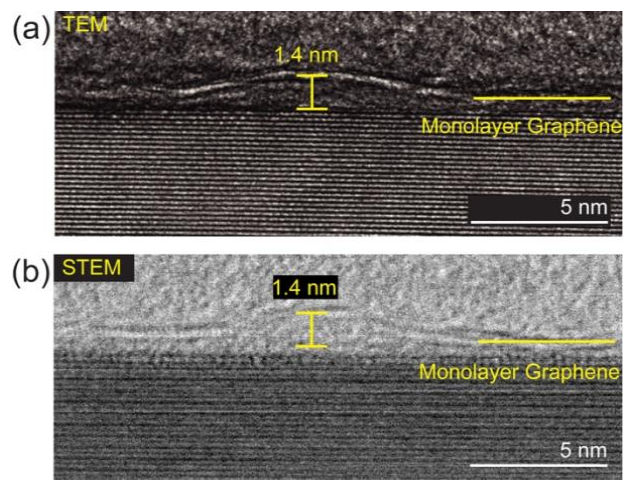

**Figure S13.** TEM (a) and STEM (b) images of a graphene wrinkle on Gr/STO. Wrinkle height is 1.4 nm, consistent with AFM profiles.

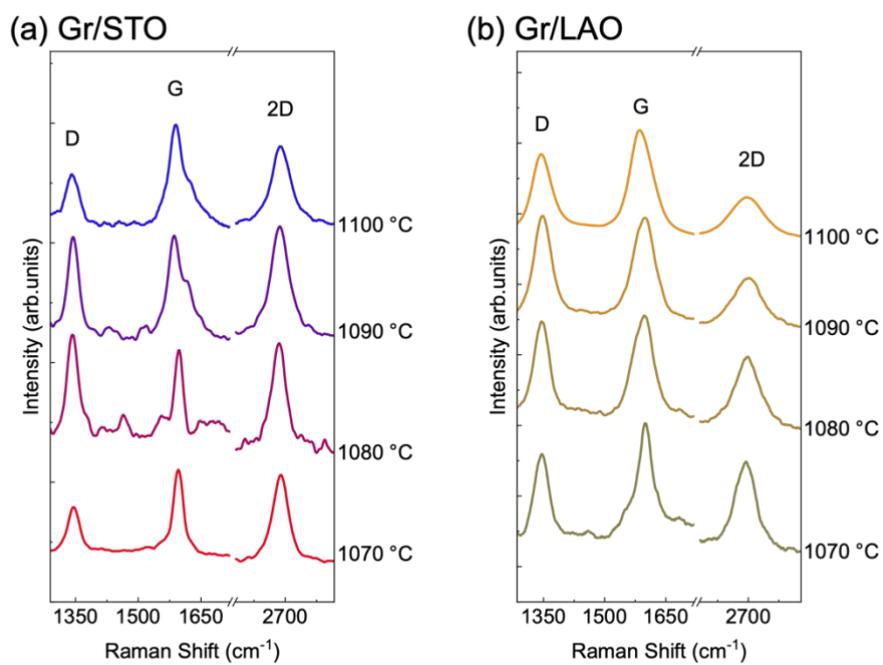

**Figure S14.** Raman spectra of HHs grown at different temperatures. (a) Gr/STO and (b) Gr/LAO with temperature labels.

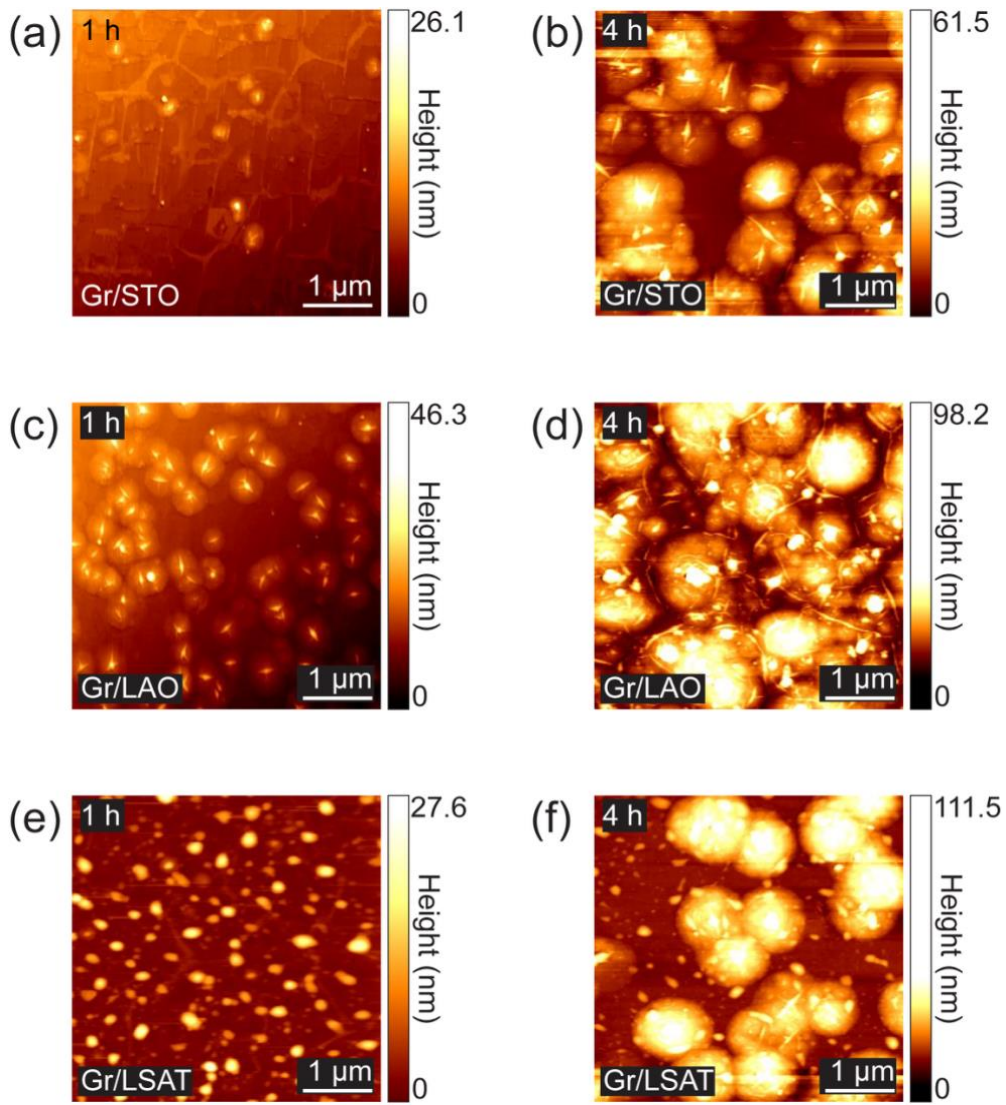

**Figure S15.** AFM of HHs grown at 1100 °C. Left column (a, c, e): 1 h growth; right column (b, d, f): 4 h growth for Gr/STO, Gr/LAO, and Gr/LSAT, respectively.

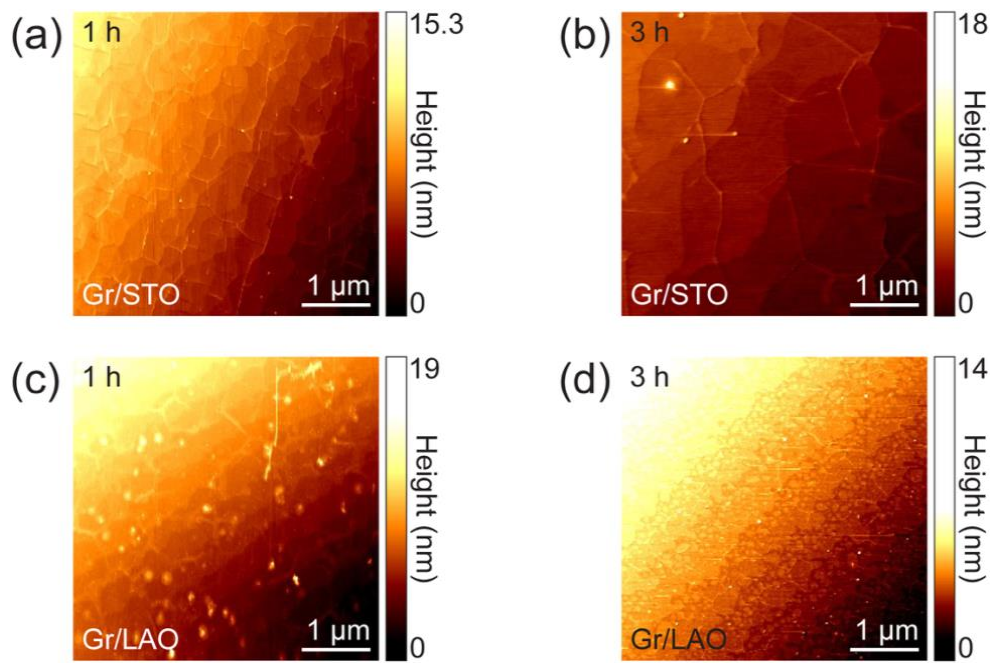

**Figure S16.** AFM of HHs grown at optimal 1070 °C. (a, c): 1h; (b, d): 3h. (a, b) Gr/STO; (c, d) Gr/LAO.

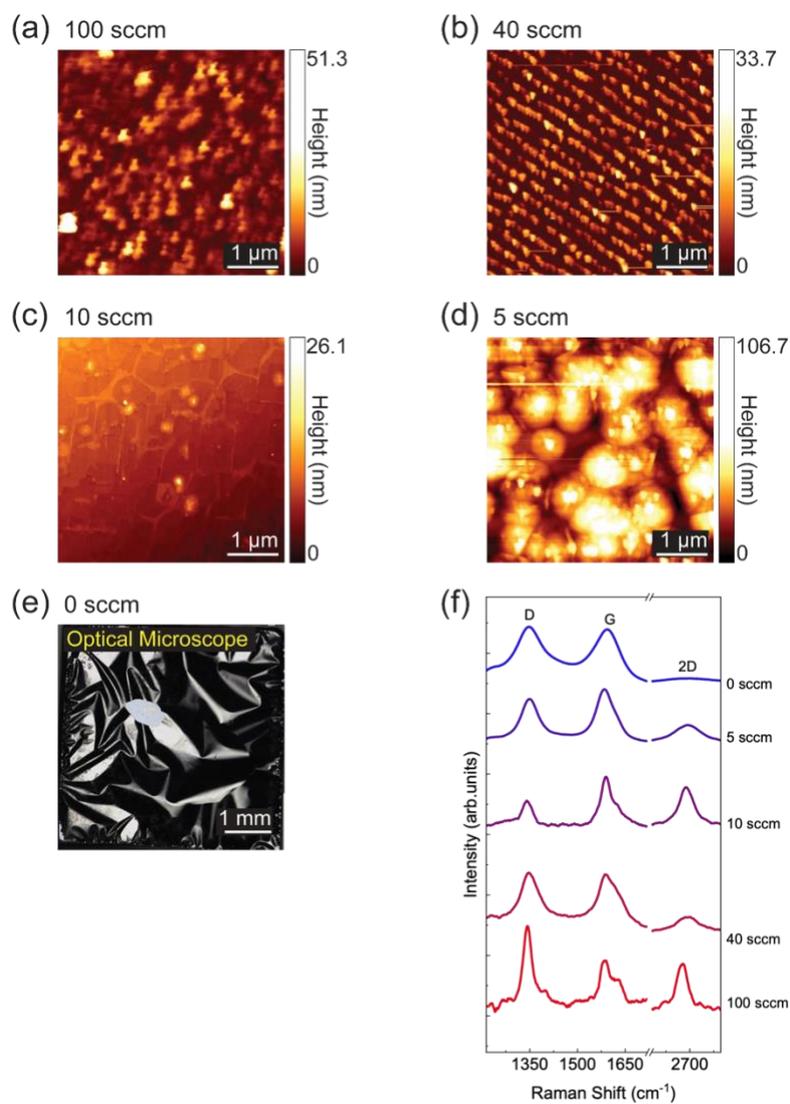

**Figure S17.** Effect of H$_2$ flow rate on Gr/STO. (a – d) AFM images at 100, 40, 10, 5, and 0 sccm. (e) Optical images showing full carbon foil coverage at 0 sccm. (f) Corresponding Raman spectra.

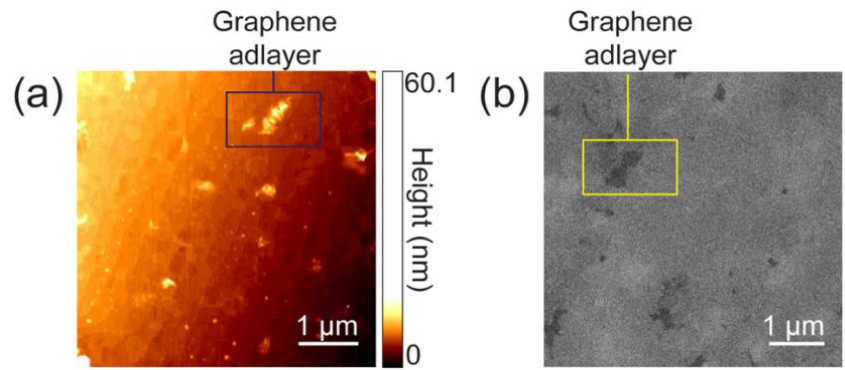

**Figure S18.** Gr/STO grown at 16 sccm CH₄. (a) AFM and (b) SEM images showing adlayers (white and dark spots).

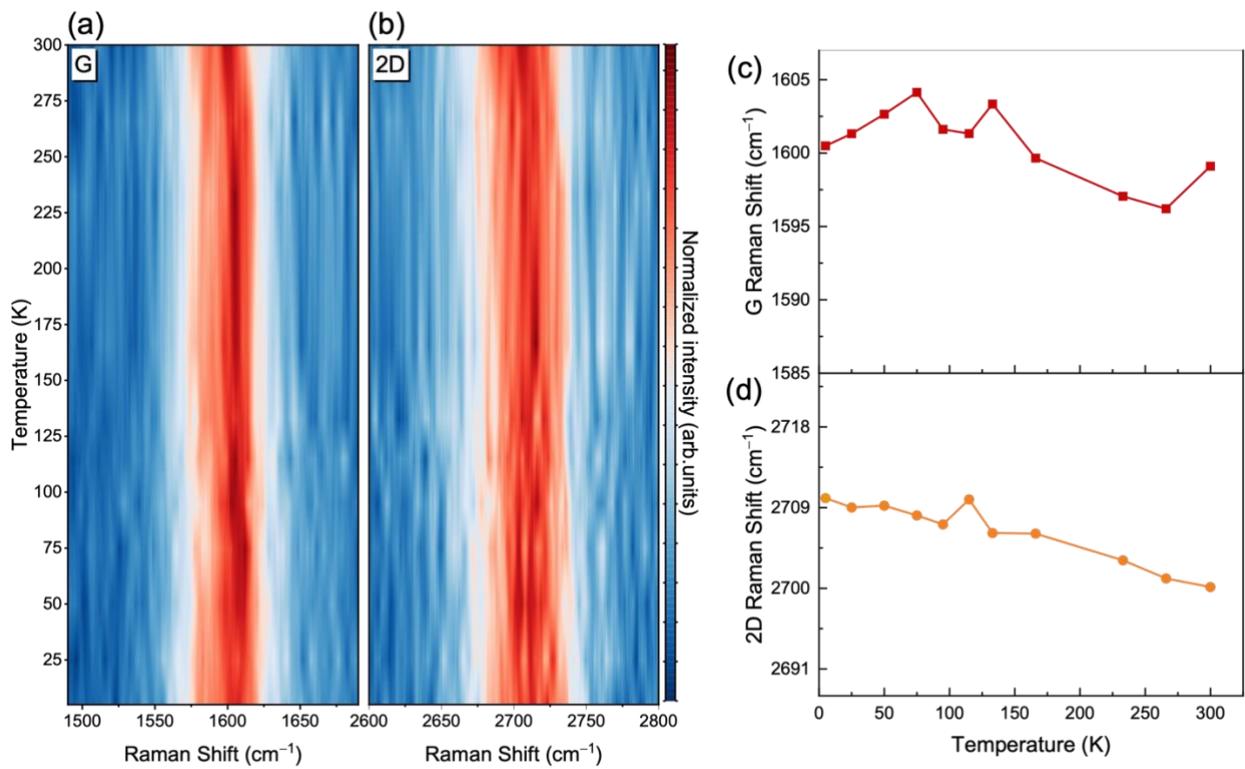

**Figure S19.** Low temperature Raman spectra of Gr/LAO HH. (a, b) G and 2D peak contour plots. (c, d) Peak positions.

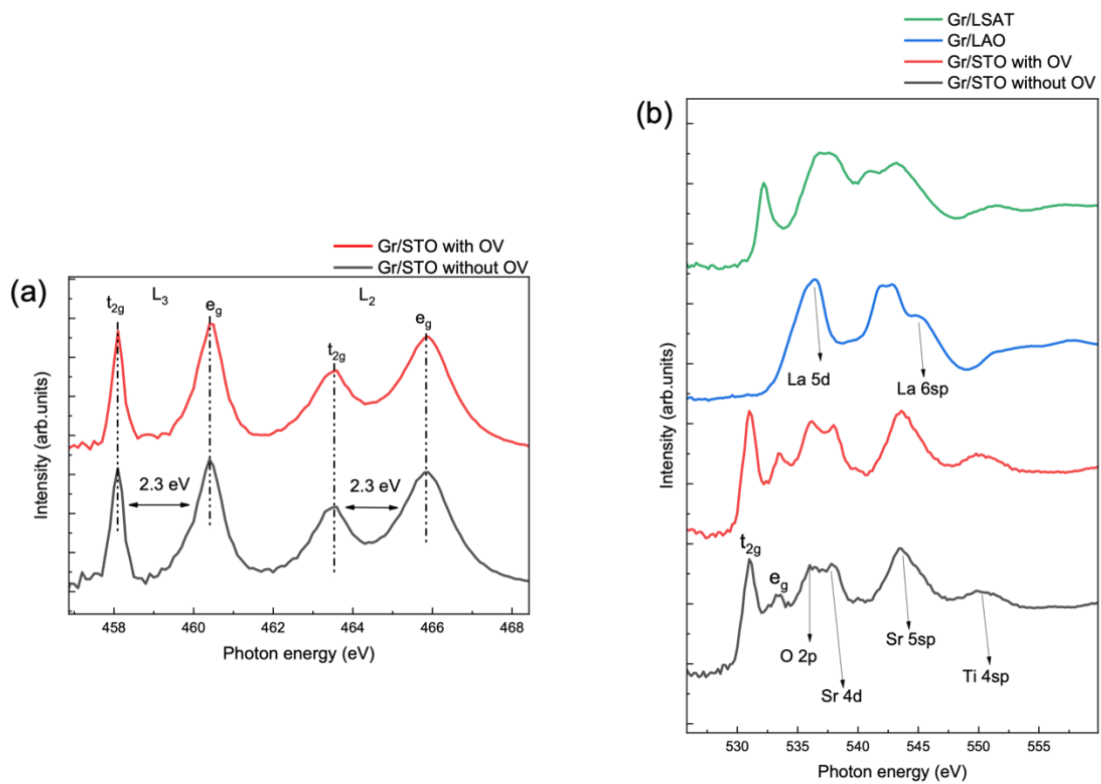

**Figure S20.** NEXAFS spectra of the Ti *L*-edge and O *K*-edge for various samples. (a) Ti *L*-edge spectra comparing the as-grown Gr/STO sample (red) with a control Gr/STO sample annealed in oxygen to restore stoichiometry (black). (b) O *K*-edge spectra for Gr/LSAT (green), Gr/LAO (blue). The Ti *L*-edge spectra for both as-grown and annealed Gr/STO samples are virtually identical and lack any distinct features corresponding to $Ti^{3+}$ species. The O *K*-edge spectra also show highly similar features between STO and STO OV sample.